\documentclass[letterpaper,english,reprint, aps]{revtex4-1}

\usepackage[T1]{fontenc}
\usepackage[latin9]{inputenc}
\usepackage{booktabs}
\usepackage{textcomp}
\usepackage{amsmath}
\usepackage{amssymb}
\usepackage{graphicx}

\usepackage[colorlinks,
            linkcolor=blue,
            anchorcolor=blue,
            citecolor=blue,
            urlcolor=blue
            ]{hyperref}

\makeatletter

\pdfpageheight\paperheight
\pdfpagewidth\paperwidth

\makeatother

\usepackage{babel}

\begin{document}

\title{BaHgSn: A Dirac semimetal with surface hourglass fermions}

\author{Tan Zhang$^{1,2}$}
\author{Zhihai Cui$^{1,2}$}
\author{Zhijun Wang$^{1,2}$}
\author{Hongming Weng$^{1,2,3,4}$}
\altaffiliation {hmweng@iphy.ac.cn}
\author{Zhong Fang$^{1,2}$}

\affiliation{${^1}$Beijing National Research Center for Condensed Matter Physics, and Institute of Physics, Chinese Academy of Sciences, Beijing 100190, China}
\affiliation{${^2}$School of Physical Sciences, University of Chinese Academy of Sciences, Beijing 100049, China}

\affiliation{${^3}$Songshan Lake Materials Laboratory , Dongguan, Guangdong 523808, China}
\affiliation{${^4}$CAS Center for Excellence in Topological Quantum Computation, University of Chinese Academy of Sciences, Beijing 100190, China}

\begin{abstract}
We proposed that BaHgSn is a Dirac semimetal (DSM) which can host hourglass-like surface states (HSSs) as protected by nonsymmorphic glide symmetry. Compared to KHgSb, an isostructural topological crystalline insulator with the same HSSs, BaHgSn has an additional band inversion at $\Gamma$ point. This band inversion is induced by the stronger interlayer coupling among Hg-Sn honeycomb layers than that among Hg-Sb-layers in KHgSb, which leads to bulk Dirac nodes in BaHgSn along the layer stacking direction $\Gamma$-$A$. In addition, the mirror Chern number $C_{i}$ protected by the mirror plane $\overline{M}_{z}$ ($k_z$=0) changes from 2 in KHgSb to 3 in BaHgSn. Therefore, when a compressive uniaxial strain is applied along the $y$ axis to break the rotation symmetry protecting the DSM state, BaHgSn becomes a strong topological insulator with $Z_{2}$ indices of $(1;000)$ and the topological surface Dirac cone co-exists with HSSs on the (010) surface. The Wilson-loop spectra have been calculated to verify these topological features. The calculated surface states, the Fermi surfaces and their quasiparticle interference patterns are ready to be compared with experimental measurements.
\end{abstract}
\maketitle

\section{INTRODUCTION\label{sec:1}}

Topological quantum states and the topological materials hosting them have been extensively studied within last decade \cite{Hasan2010,Qi2011, AdvancesinPhysics64.2015, Chiu2016, Weng2014,Weng2016, 10.1093nsrnwx066, RevModPhys.90.015001}. One of the most important and exciting finding is the realization of several relativistic quasiparticles as emergent effects at the boundaries where topological phase transition happens. For example, the two-dimensional (2D) massless Dirac fermions have been realized on the surface of three-dimensional (3D) topological insulators (TIs)~\cite{Hasan2010,Qi2011} and the Weyl fermions have been realized in 3D Weyl semimetals (WSMs), which can be viewed as the boundary state of a four-dimensional quantum Hall insulators \cite{RevModPhys.90.015001, Weng2016}. Up to now, there have been various quasiparticles theoretically proposed with and without their counterparts in field theory. Several of them have even been experimentally observed in some topological materials, such as the Dirac fermions in Dirac semimetals (DSMs)  \cite{Wang2012, PhysRevB.88.125427}, Majorana fermions in topological superconductors \cite{Qi2011,Zhang182}, Weyl fermions in WSMs \cite{PhysRevX.5.011029, PhysRevX.5.031013, Xu613, Yang2015, Wang2018, Liu2018}, hourglass fermions in topological crystalline insulators \cite{Wang2016, Mae1602415} and spin-1 fermions in triply degenerate nodal-point semimetals \cite{Bradlynaaf5037, PhysRevB.93.241202, Nature546, PhysRevX.6.031003, 10.1093nsrnwx066}.

The hourglass fermions were proposed and observed on the surface of a typical topological crystalline insulator (TCI) KHgSb in nonsymmorphic space group $P6_{3}/mmc$ ($D_{6h}^{4}$) \cite{Wang2016, Mae1602415}. 
It has symmetry-based indicator $\mathbb{Z}_{12}=8$ \cite{Zhang2019,Bernevig2019nature,Tang2019nature} and nontrivial topological invariants, namely hourglass invariant $\delta_h=1$ in glide plane $\overline{M}_{x}$ and mirror Chern number (mCN) $C_i=2$ in $\overline{M}_z$ $(k_z=0)$ plane. These topological invariants determine that it has hourglass-like surface states (HSSs) and Dirac cone like ones protected by these symmetries on a surface preserving them~\cite{Song2018}.
There have been many compounds of this crystal structure proposed to be topological materials. For example, XYBi (X = Ba, Eu; Y = Cu, Ag and Au) family \cite{Du2015,QIN2019218} and XAuTe (X = K, Na, Rb) family \cite{Sun2016, QIN2019218} were proposed to realize DSMs or WSMs after proper tuning.
It is noted that NaAuTe was predicted to be a DSM~\cite{QIN2019218} since it has band crossing induced by the stronger bonding-antibonding splitting than KHgSb, which also leads to mCN $C_{i}=3$ for the mirror plane with $k_z=0$.
The similar way to achieving this band crossing has been proposed by applying compressive pressure along the $c$ axis in KHgSb family compounds~\cite{npj2057}. For example, KHgBi becomes a DSM under 11.5\% compression of $c$ lattice constant although such large compression is not plausible. 
Here, we propose that BaHgSn is another DSM in the same crystal structure with the same physical mechanism. Therefore, it can host Dirac fermions in bulk and hourglass fermions on the (010) surface. When the rotation symmetry $C_{3z}$ protecting Dirac nodes is broken, it becomes a strong topological insulator \cite{Wang2012}. Combining theoretical analysis, first-principles calculations and Wilson loop calculations, we can well understand the resultant nontrivial surface states, including the HSSs and the zigzag like ones protected by mCN $C_{i}=3$ proposed in Ref.~\onlinecite{Wang2016} but not been found in real materials.
These novel surface states can be identified by the calculations of surface states, the Fermi surfaces, and their  quasiparticle interference (QPI) patterns. These phenomena can be readily measured by angle-resolved photoemission spectroscopy (ARPES) \cite{RevModPhys.82.3045,Hsien2009,PhysRevLett.104.016401} and scanning tunneling microscope (STM) \cite{Inoue1184,PhysRevB.93.041109,Li2019,Yuaneaaw9485}.

\section{CRYSTAL STRUCTURE AND METHODOLOGY\label{sec:2}}

\begin{figure}
\includegraphics[width=1\columnwidth]{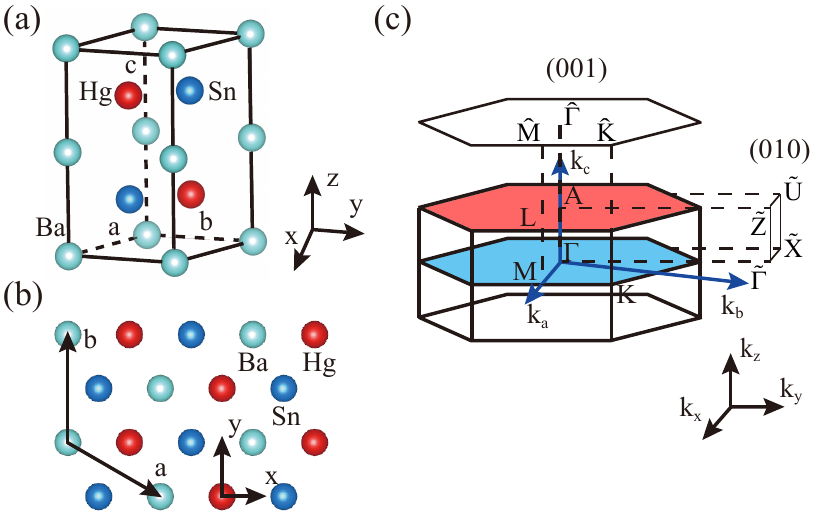}
\caption{(Color online) Crystal structure and Brillouin zone (BZ) of BaHgSn.
(a) The crystal structure of BaHgSn primitive cell. $a$, $b$ and $c$ are lattice vectors. The Ba ion (cyan) is at the spatial origin, and the Hg (red) and Sn (blue) ions form two AA$^\prime$ stacking honeycomb layers. 
(b) Top-down view of the crystal structure. 
(c) Bulk BZ of BaHgSn. The blue and red planes are two mirror planes of $\overline{M}_{z}$ with $k_z$=0 and $k_z$=$\pi$, respectively. The (010) and (001) surface BZs are plotted. 
}
\label{fig:1}
\end{figure}

The crystal structure of BaHgSn \cite{Merlo1993,Vogel1980} is the same as KHgSb \cite{Wang2016} as illustrated in Fig. \ref{fig:1}. 
They belong to the same nonsymmorphic space group $D_{6h}^{4}$ $(P6_{3}/mmc)$, and the symmetry operators include: an inversion $P$, a screw rotation $\overline{C}_{6z}$, 
mirror operations $M_{y}$, mirror operations $\overline{M}_{z}=t(c\overrightarrow{z}/2)M_{z}$ and glide mirror operations $\overline{M}_{x}=t(c\overrightarrow{z}/2)M_{x}$, where $t(c\overrightarrow{z}/2)$ is a $c/2$ translation along $z$ axis. The experimental lattice parameters of BaHgSn are $a=b=5.012$ \AA~and $c=9.715$ \AA. 
The Ba, Hg and Sn ions are at the same Wyckoff positions as K, Hg and Sb ions in KHgSb. The Hg, Sn ions form AA$^\prime$ stacking honeycomb layers intercalated by trigonal layers of Ba ions.

We perform first-principles calculations of BaHgSn based on the projector augmented plane-wave (PAW) method \cite{Blochl1994,Kresse1999}.
The Perdew-Burke-Ernzerhof (PBE) exchange-correlation functional is used under  generalized gradient approximation (GGA) \cite{Perdew1992}. 
The cutoff energy is 450 eV, and the k-point sampling grid is 12\texttimes{}12\texttimes{}4 $\Gamma$-centered mesh.
The hybrid functional calculations in HSE~\cite{Heyd2003,Heyd2006} scheme are further applied
to improve the shortcomings of GGA. 
To study the (010) and (001) surface states, the surface Green\textquoteright s function technique is used to simulate the semi-infinite system by constructing the Wannier tight-binding Hamiltonian \cite{PhysRevB.47.1651,Wieder246,WU2017}, which are generated for $s$ orbitals of Hg and $p$ orbitals of Sn. 
The Fermi surfaces and spin-dependent scattering probability (SSP) which can simulate the QPI patterns detected by the STM are calculated according to the formula in Ref. \onlinecite{Inoue1184} on the basis of Wannier Hamiltonian. Because STM is very sensitive to the surface features, the principle layers containing four unit cells is used to construct the surface Green\textquoteright s function but only the contributions in the outermost one unit cell are weighted. The Wilson-loop spectra are also computed to verify the topological features of surface states~\cite{PhysRevB.47.1651,Wieder246}.

\section{RESULTS AND DISCUSSIONS\label{sec:3}}

\subsection{Electronic structures: bulk analysis}

\begin{figure}
\includegraphics[width=1\columnwidth]{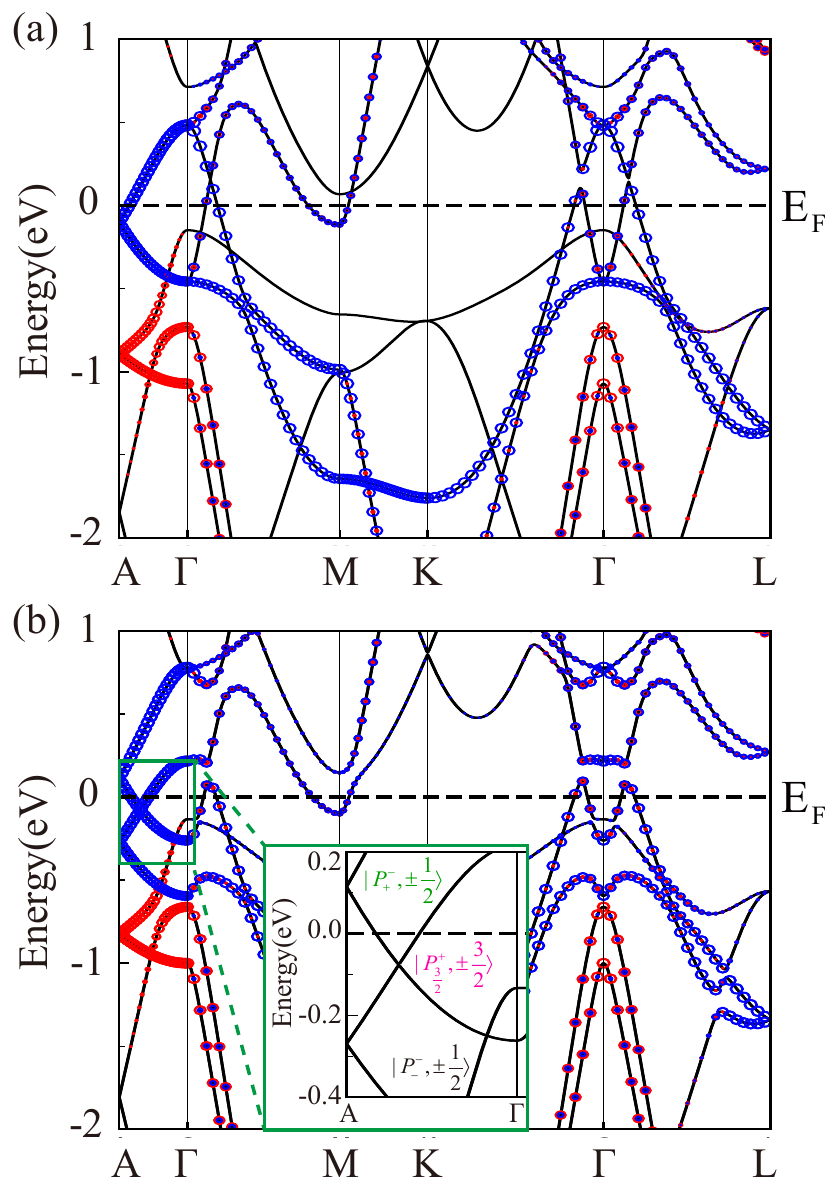}
\caption{(Color online) The calculated bands of BaHgSn (a) without and (b) with spin-orbit coupling. The blue circles indicate the projection weight of Sn $p_x$ and $p_y$ orbitals, and the red circles indicate that of the Hg $s$ orbitals. (c) The bulk Dirac cone and orbital characters of the wave functions at $\Gamma$ point.
\label{fig:2}}
\end{figure}

The electronic structures are shown in Fig. \ref{fig:2}. They are quite similar with KHgSb. The bands from Hg $s$ orbitals are lower than those from Sn (or Sb in KHgSb) $p$ orbitals. In the case without spin-orbit coupling (SOC), the valence and conduction bands at $A$ point around Fermi energy are degenerate since they are from Sn $p_{x,y}$ orbitals in the two Hg-Sn honeycomb layers in one unit cell. The interlayer coupling leads to the bonding-antiboning splitting along $\Gamma$-$A$ path. 
When SOC is considered as shown in Fig. \ref{fig:2}(b), these bands further split into sub-bands with different angular momenta. It leads to a band inversion and a pair of Dirac nodes along the $\Gamma$-$A$ path. The Dirac nodes are around -0.08 eV below the Fermi energy. The band inversion is about 0.48 eV at $\Gamma$ point, and further HSE calculation corrects it to be about 0.44 eV. Therefore, this band inversion is most plausible in realistic material.

Due to the existence of inversion symmetry, we can use bonding and anti-bonding states with definite parity to describe the low-energy Hamiltonian around Fermi energy. 
The bonding and anti-bonding states of $p$ orbitals are $|P_{\alpha}^{\pm}\rangle=\frac{1}{\sqrt{2}}\left(|\textrm{Sn1},p_{\alpha}\rangle\mp|\textrm{Sn2},p_{\alpha}\rangle\right)$, where Sn1 and Sn2 are equivalent Sn atoms related with inversion symmetry, the superscripts $\pm$ indicate the parity (the bonding or anti-bonding states), and $\alpha$ is $x,y,z$. 
When SOC is taken into account, the spin and orbital angular momentum are coupled, and the eigenstates have definite total angular momentum, written as $|P_{\frac{3}{2}}^{\pm}, \pm\frac{3}{2}\rangle$, $|P_{\frac{3}{2}}^{\pm},\pm\frac{1}{2}\rangle$, $|P_{\frac{1}{2}}^{\pm},\pm\frac{1}{2}\rangle$, where the subscript indicates the total angular momentum $J$, the $\pm\frac{3}{2}$, $\pm\frac{1}{2}$ indicate the eigenvalues of $J_{z}$.
The heavy-hole state $|P_{\frac{3}{2}}^{\pm},\pm\frac{3}{2}\rangle$ and light-hole states $|P_{\frac{3}{2}}^{\pm},\pm\frac{1}{2}\rangle$  are no longer degenerated at $\Gamma$ point. 
The light-hole states $|P_{\frac{3}{2}}^{\pm},\pm\frac{1}{2}\rangle$ and states $|P_{\frac{1}{2}}^{\pm},\pm\frac{1}{2}\rangle$ will mix further to form the new eigenstates: $|P_{+}^{-},\pm\frac{1}{2}\rangle$, $|P_{+}^{+},\pm\frac{5}{2}\rangle$, $|P_{-}^{-},\pm\frac{1}{2}\rangle$ and $|P_{-}^{+},\pm\frac{5}{2}\rangle$ \cite{Liu2010}.
The states included in the band inversion and crossing points are states $|P_{+}^{-},\pm\frac{1}{2}\rangle$ and $|P_{\frac{3}{2}}^{+},\pm\frac{3}{2}\rangle$ as shown in Fig. \ref{fig:2}(c). 
Because of inversion and time-reversal symmetries, each band has Kramer degeneracy. 
The crossing points are fourfold degenerate Dirac nodes, locating at $(0,0,k_z=\pm0.35\times\frac{\pi}{c})$. They are protected by $C_{3z}$ rotation symmetry \cite{Wang2012}.

\begin{figure*}
\includegraphics[width=2\columnwidth]{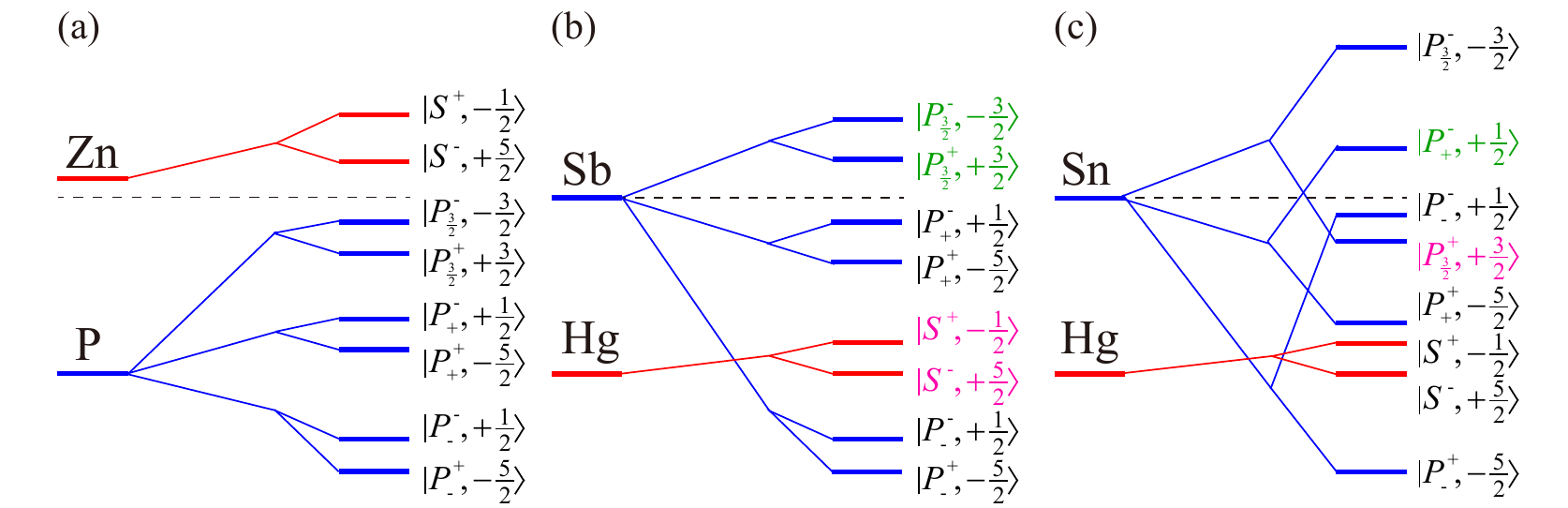}
\caption{(Color online) The schematic diagrams show the evolution from the atomic $p$ orbitals (blue line) and $s$ orbitals (red line) to the conduction and valence states at $\Gamma$ for (a) KZnP, (b) KHgSb and (c) BaHgSn. The evolution includes the effects of bonding-antibonding splitting and spin-orbit coupling. 
Only the states having $+i$ eigenvalue of $\overline{M}_{z}$ ($k_z$=0) are marked here. The inverted occupied orbitals are marked in pink and the inverted unoccupied ones are in green. The black dashed line represents the Fermi energy.
\label{fig:3}}
\end{figure*}

We compare the band structures of KZnP, KHgSb and BaHgSn since they are iso-structural while have different topological quantum states. The key difference is the energy order of the bands around Fermi level at $\Gamma$,  including $s$ bands from Zn or Hg and $p$ bands from of P, Sb and Sn.
The schematic diagrams for the evolution of these bands are shown in Fig.~\ref{fig:3}.
KZnP is a trivial insulator and is used as a reference system having the normal order of bands, namely the cation's $s$ bands are higher than the anion's $p$ bands in energy. 
KHgSb is a well known TCI and it has two Hg $s$ orbitals inverting with two Sb $p$ orbitals~\cite{Wang2016}.
BaHgSn has one more band inversion among Sn $p$ bands at $\Gamma$ point than KHgSb since the interlay coupling in BaHgSn is stronger than the strength of SOC. 
Sn and Sb atoms have the similar SOC strength as estimated from the splitting between $|P_{+}^{-},\pm\frac{1}{2}\rangle$ and $|P_{\frac{3}{2}}^{-},\pm\frac{3}{2}\rangle$), but BaHgSn has larger energy splitting in the interlayer bonding and antibonding states than KHgSb. This can be seen from the splitting between $|P_{+}^{-},\pm\frac{1}{2}\rangle$ and $|P_{+}^{+},\pm\frac{5}{2}\rangle$, or that between $|P_{\frac{3}{2}}^{+},\pm\frac{3}{2}\rangle$ and $|P_{\frac{3}{2}}^{-},\pm\frac{3}{2}\rangle$. 
It is due to the smaller $c$ lattice constant of BaHgSn (9.71 \AA) than that of KHgSb (10.22 \AA). 
The weak interlayer bonding-antibonding splitting in KHgSb results in a TCI, but the stronger splitting in BaHgSn makes it a DSM. The difference in interlay coupling strength comes from the smaller lattice constant $c$ of BaHgSn than that of KHgSb by about 0.51~\AA. This may be caused by the stronger attractive interaction of the $\pm 2$ valence of Ba ions and Hg-Sn honeycomb layers than that of $\pm 1$ valence of K ions and Hg-Sb layers, although the ionic radius of K$^{+1}$ and Ba$^{+2}$ are nearly the same~\cite{Shannon12967}.

The states at $\Gamma$ point are eigenstates of both inversion and $\overline{M}_{z}$.
For occupied bands, the $Z_2$ invariant can be easily determined by counting the parity of their eigenstates~\cite{Hasan2010,Qi2011}.
It is known that the mCN $C_i$ in mirror plane $\overline{M}_{z}$ with $k_z$=0 can be obtained by calculating the eigenvalues of rotation symmetry in subspace of eigenstates with either +$i$ or $-i$ mirror eigenvalue~\cite{PhysRevB.86.115112,Wang2016,QIN2019218}, as following:
\begin{equation}
\label{eq:e1}
\begin{aligned}
e^{-i\frac{\pi}{3}C}&=&\prod_{i\in{occ.}}(-1)^{F}\eta_i(\Gamma)\theta_i(K)\zeta_i(M), 
\end{aligned}
\end{equation}
where $F=0\ (1)$ for integer-spin (half-integer-spin) system, and $\eta_i(\mathbf k)$, $\theta_i(\mathbf k)$ and $\zeta_i(\mathbf k)$ represent the eigenvalues of $C_{6z}$, $C_{3z}$ and $C_{2z}$ rotation operators at the corresponding rotational invariant momenta, respectively.

We start from the trivial insulator KZnP in Fig. \ref{fig:3}(a). Its $Z_2$ invariant and mCN for $k_z$=0 plane are both zero.
Compared with KZnP, KHgSb has two band inversions as shown in Fig. \ref{fig:3}(b). We only consider the $+i$ mirror eigenvalue subspace. 
The newly occupied orbitals are $|S^{+},-\frac{1}{2}\rangle$ and $|S^{-},+\frac{5}{2}\rangle$, and the newly unoccupied orbitals are $|P_{\frac{3}{2}}^{-},-\frac{3}{2}\rangle$ and $|P_{\frac{3}{2}}^{+},+\frac{3}{2}\rangle$.
Obviously, the change in total parity $\Delta \delta=0$ and that in rotational eigenvalue is $\Delta J_z=2$, which means that $Z_2=0$ and the mCN of KHgSb is calculated from 
\begin{equation}
\label{eq:e2}
\begin{aligned}
e^{-i\frac{\pi}{3}C_{i}}=e^{-i\frac{\pi}{3}2}.
\end{aligned}
\end{equation}
The mCN of KHgSb is determined as $C_{i}=2 \mod 6$.
Compared with KHgSb, BaHgSn has one additional band inversion for $+i$ mirror eigenvalue subspace in Fig. \ref{fig:3}(c).
The newly occupied orbital is $|P_{\frac{3}{2}}^{+},+\frac{3}{2}\rangle$, and the newly unoccupied orbital is $|P_{+}^{-},+\frac{1}{2}\rangle$. 
Thus, the change of parity $\Delta \delta=1$ and that of $\Delta J_z=1$, indicating that the $Z_2=1$ and mCN $C_{i}=3$ from
\begin{equation}
\label{eq:e3}
\begin{aligned}
e^{-i\frac{\pi}{3}C_{i}}=e^{-i\frac{\pi}{3}(2+1)}.
\end{aligned}
\end{equation}
The odd mCN of $k_z$=0 plane means odd $Z_2$, which is consistent with $\Delta \delta=1$.
If 5\% compression strain is applied along the $y$ axis in BaHgSn to break the $C_{3z}$ rotation symmetry~\cite{Wang2012}, the Dirac nodes on $\Gamma$-$A$ will be gapped out by about $9.2$ meV, and BaHgSn becomes a strong TI with $Z_{2}$ indices of $(1;000)$.

\subsection{Surface analysis}

\begin{figure*}
\includegraphics[width=2\columnwidth]{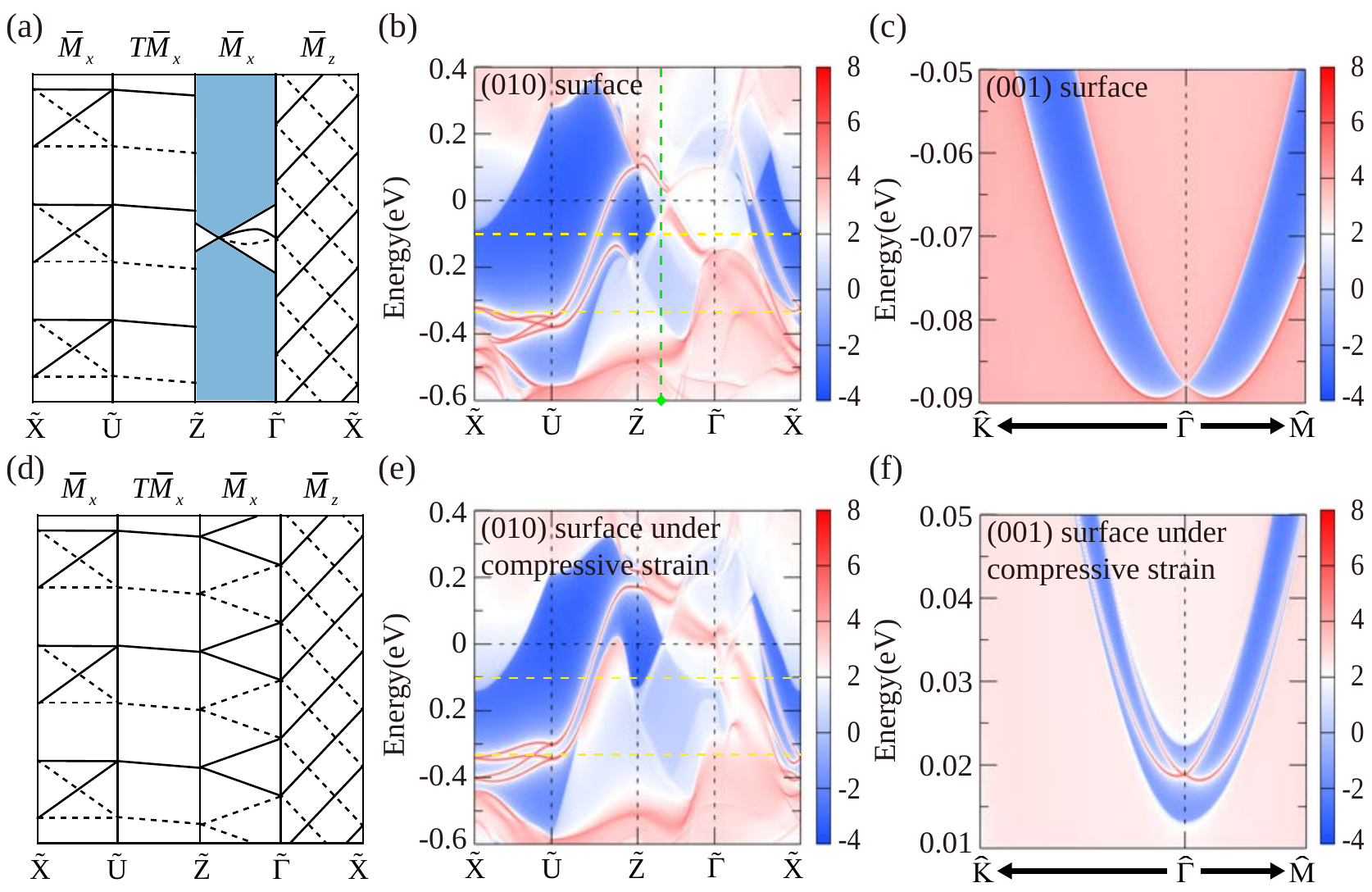}
\caption{(Color online) (a) The schematic diagram of surface bands for (010) surface. (b) The calculated (010) surface states. The light green dashed line and dot indicate the projection position of the Dirac node. (c) the calculated (001) surface states. (d), (e) and (f) are  same as (a), (c), and (e), respectively, except that the 5\% compressive strain is applied along $y$ axis. The horizontal dashed lines in (b) and (e) indicate the Fermi energy at 0.0 (charge neutral level), -0.1 and -0.33 eV, respectively. 
\label{fig:4}}
\end{figure*}

After understanding the band topology of KZnP, KHgSb and BaHgSn, we turn to their topological surface states.
For KZnP, it might have trivial surface states.
For KHgSb, on its (010) surface it has HSSs protected by glide symmetry $\overline{M}_{x}$ and surface states due to $C_i$=2 as protected by mirror symmetry $\overline{M}_{z}$ ($k_z$=0)~\cite{Wang2016,Song2018}. For BaHgSn, we have calculated the surface states for (010) and (001) surfaces as shown in Fig.~\ref{fig:4}. 
Due to additional band inversion and different topological invariant, BaHgSn is expected to have some different features in surface states other than KHgSb.

For (010) surface, the mirror $\overline{M}_{z}$ and glide mirror $\overline{M}_{x}$ are preserved.
To explain the nontrivial surface states, we consider each high-symmetrical line in turn.
Along the paths of $\tilde{X}\tilde{U}$ and $\tilde{Z}\tilde{\Gamma}$, $\overline{M}_{x}$ is preserved. 
Because of time-reversal symmetry $T$, the eigenstates with ${+i,-i}$ eigenvalue of $\overline{M}_{x}$ are degenerated at $\tilde{X}$ and $\tilde{\Gamma}$ $(k_{z}=0)$, and those of ${+1,+1}$ or ${-1,-1}$ are degenerated at $\tilde{U}$ and $\tilde{Z}$ $(k_{z}=\pi/c)$.
These constraints imply two topologically distinct connectivities for the surface states, one of which is nontrivial zigzag connectivity across the band gap \cite{Hsieh2008,Koenig2007,Fu2007,Moore2007,Roy2009}, the other one is hourglass-shaped dispersion. 
We can distinct these two type surface states by $Z_2$ invariant.
For KHgSb, it has trivial $Z_{2}$ indices $(0;000)$, so that the HSSs exist along both $\tilde{X}\tilde{U}$ and $\tilde{Z}\tilde{\Gamma}$ paths. 
For BaHgSn, the similar HSSs appear only along $\tilde{X}\tilde{U}$ path. Along $\tilde{Z}\tilde{\Gamma}$ path, the surface states are not clear since the bulk Dirac node is projected onto this path as shown in Fig.~\ref{fig:4}(a). However, if the bulk Dirac node is gapped by breaking $C_{3z}$ rotation symmetry, the surface states along $\tilde{Z}\tilde{\Gamma}$ should be zigzag as shown in Fig.~\ref{fig:4}(d) since the $Z_2$ indices are (1;000) as a strong TI.
The calculated (010) surface states are shown in Fig.~\ref{fig:4}(b). They are hourglass-shaped dispersions along $\tilde{X}\tilde{U}$ path. Because of the overlap of bulk bands, the projection of the bulk Dirac cone and surface states are not clear along $\tilde{Z}\tilde{\Gamma}$ path.

Along $\tilde{U}\tilde{Z}$ $(k_{z}=\pi/c)$ path, the symmetry is $T\overline{M}_{x}$, which results in a Kramers-like degeneracy. The surface states of BaHgSn are similar as those of KHgSb as shown in Figs.~\ref{fig:4}(a) and (b).

Along $\tilde{\Gamma}\tilde{X}$ $(k_{z}=0)$ path, the surface states are determined by the mCN $C_{i}$ of $\overline{M}_{z}$ with $k_z=0$ \cite{Teo2008,PhysRevX.6.021008}. 
For KHgSb, there are two surface bands protected by $C_{i}=2$ crossing the Fermi level as shown in Ref.~\onlinecite{Wang2016}. 
But for BaHgSn, there are three surface bands protected by $C_{i}=3$ crossing the Fermi level as shown in the Figs. \ref{fig:4}(a) and (d), though such surface states are buried by the projection of bulk bands in Fig. \ref{fig:4}(b).

For (001) surface, the surface states of KHgSb are trivial. But for BaHgSn, the projection of bulk Dirac cone exists at $\hat{\Gamma}$ point as shown in Fig.~\ref{fig:4}(c). When 5\% compressive strain is applied along the $y$ axis, the gapless Dirac points open gaps and the surface Dirac cone appears in Fig.~\ref{fig:4}(f).

\subsection{Quasiparticle interference on (010) surface}

\begin{figure*}
\includegraphics[width=2\columnwidth]{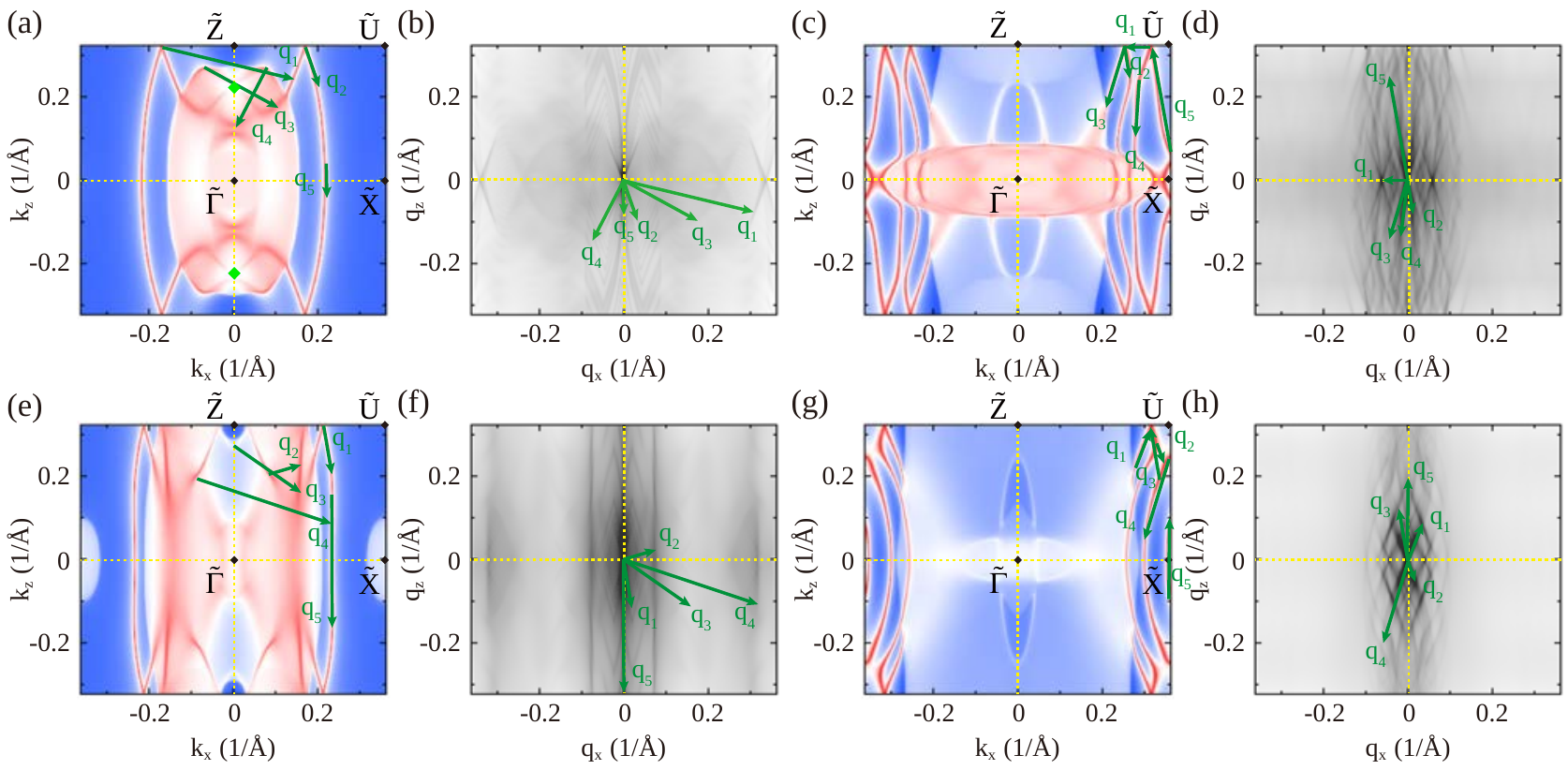}
\caption{(Color online) (a) The Fermi surface of (010) surface with Fermi energy $E_F$= -0.1 eV in the first surface BZ. The green arrows indicate the prominent scattering vectors. The light green dots indicate the projected positions of the Dirac nodes. (b) The corresponding QPI pattern simulated for (010) surface with $E_F$= -0.1 eV.
(c) and (d) The Fermi surface and QPI pattern of (010) surface with $E_F$= -0.33 eV. 
(e), (f), (g) and (h) The Fermi surfaces and QPI patterns of (010) surface with $E_F$= -0.1 eV and -0.33 eV when 5\% compressive strain is applied along $y$ axis. 
}
\label{fig:5}
\end{figure*}

The Fermi surface of (010) surface with Fermi energy $E_F$ set at -0.1 eV is presented in Fig. \ref{fig:5}(a). This makes $E_F$ very close to the energy of Dirac nodes and the projections of two Dirac nodes are indicated with dots on the $\tilde{\Gamma}\tilde{Z}$ path. It can be clearly seen that there are surface states merging into the bulk Dirac cones to form Fermi arcs similar to those in DSMs Na$_3$Bi, Cd$_3$As$_2$ and others~\cite{Wang2012,PhysRevB.88.125427,Peng:2018wf}. It is noted that there are two surface states meet each other to be doubly degenerate on the path $\tilde{U}$-$\tilde{Z}$, which is consistent with the discussion in former subsection.
The weight of scatterings are mainly determined by local density of states (LDOS), but the scatterings between surface states at $\bold k$ and $-\bold k$ points are prohibited, since the time-reversal symmetry $T$ inverses their spin directions~\cite{RevModPhys.82.3045,Hsien2009,PhysRevLett.104.016401}.
The green arrows mark the prominent scattering vectors, while the scatterings under time-reversal symmetry and mirror $\overline{M}_{z}$ symmetry are not marked. 
The corresponding QPI pattern is shown in Fig. \ref{fig:5}(b) with prominent scatterings marked as green arrows. It shows the scattering from surface states along $\tilde{\Gamma}\tilde{Z}$ path labeled as $q_4$ and crossing-shaped pattern at Bragg point.
The HSSs can be shown by Fermi surface and QPI pattern with $E_F$ at -0.33 eV in Figs. \ref{fig:5}(c) and (d). 
Two HSSs are shown along the $\tilde{X}\tilde{U}$ path, and two doubly degenerate surface states are shown along the $\tilde{U}\tilde{Z}$ path. The corresponding QPI pattern becomes ribbon-shaped along the $k_z$ axis since complicated scatterings. A crossing-shaped pattern still exist at Bragg point, and two small crossing-shaped patterns labeled as $q_1$ exist beside it.

When 5\% compressive strain is applied along $y$ axis, the bulk Dirac nodes are gapped. The Fermi surfaces and QPI patterns at -0.1 eV and -0.33 eV are shown in Figs. \ref{fig:5}(e), (f), (g) and (h). The zigzag surface states exist along the $\tilde{\Gamma}\tilde{Z}$ path at -0.1 eV and their scattering $q_3$ can be found in QPI pattern. The line-shaped QPI patterns along $k_z$ axis in Fig. \ref{fig:5}(f) are generated from scatterings of line-shaped surface states. 
The surface states at -0.33 eV become narrow when compressive strain is applied, so that their QPI pattern becomes narrow from Fig. \ref{fig:5}(d) to (h). Except that, the combination of crossing-shaped and rhombus-shaped QPI patterns emerges.

\subsection{Wilson-loop spectra on (010) surface}

\begin{figure}
\includegraphics[width=1\columnwidth]{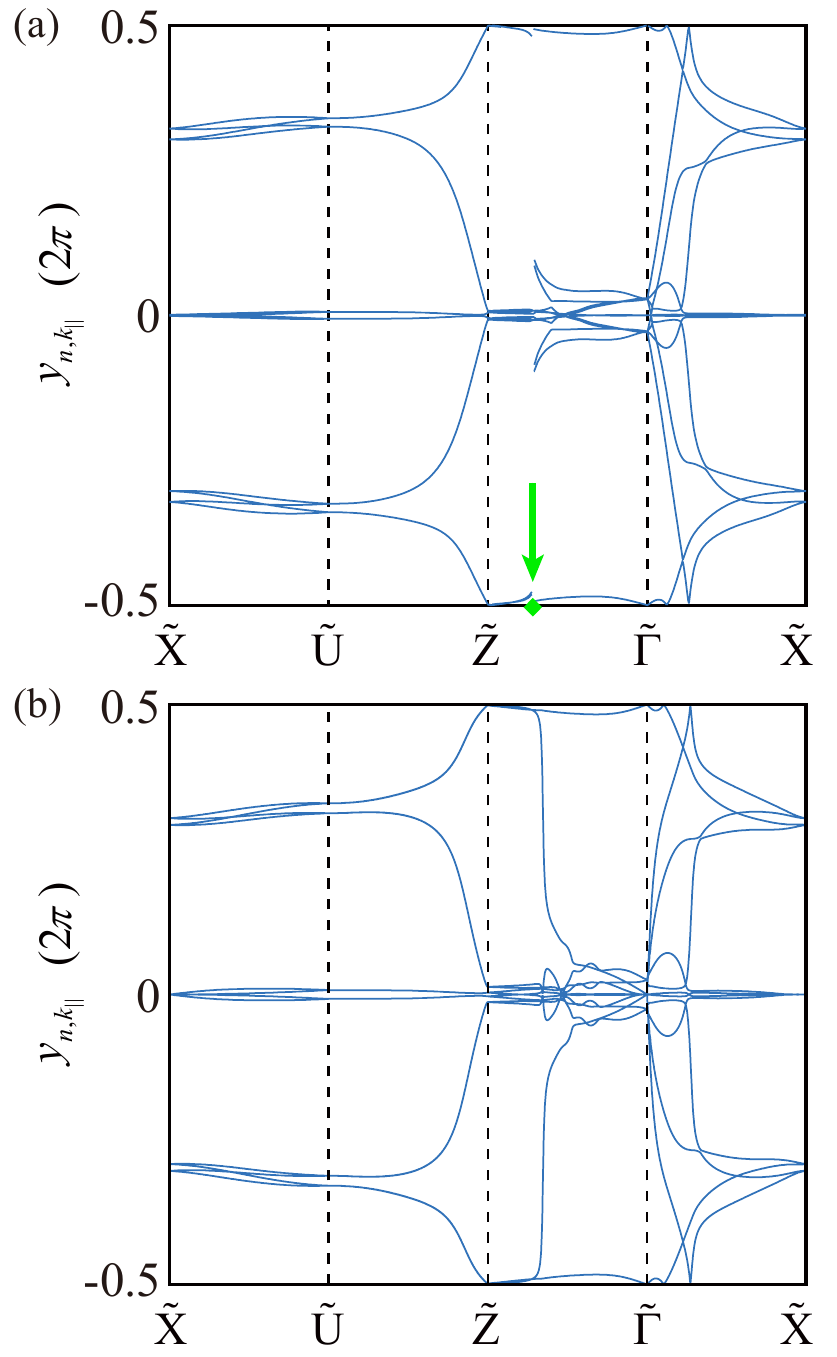}
\caption{(Color online) The Wilson-loop spectra for (010) surface of BaHgSn. 
(a) The case without compressive strain and (b) with 5\% compressive strain along the $y$ axis. The light green arrow and dot indicate the projection position of the Dirac node.} 
\label{fig:6}  
\end{figure}

The (010) surface states can also be described formally by Wilson-loop spectra based on maximally localized Wannier function for the occupied states~\cite{PhysRevB.89.155114, PhysRevB.47.1651}.
Each Wannier function is an eigenfunction of the projected position operator \cite{Wang2016}.
The eigenvalue $y_{n,\boldsymbol{k}_{\parallel}}$ is the centre of one-dimensional Wannier function along $k_y$ direction, which is a function of $\boldsymbol{k}_{\parallel}=(k_x,k_z)$ and is referred as Wilson-loop spectra \cite{PhysRevB.47.1651}.

We consider the full gapped occupied bands $\left(n_{occ}=12\right)$. The eigenvalue $y_{n,\boldsymbol{k}_{\parallel}}$ is calculated without and with 5\% compressive strain along the $y$ axis in Fig.~\ref{fig:6}. 
The spectra have the topologically equivalent features as (010) surface states~\cite{Fidkowski2011,Huang2012}: (i) the hourglass-shaped quadruplets along $\tilde{X}\tilde{U}$ path; (ii) the degenerate doublets along $\tilde{U}\tilde{Z}$ path; and (iii) the crossings protected by mCN $C_i=3$ along $\tilde{\Gamma}\tilde{X}$ path.
However, the Wilson-loop spectra are not well defined at the projection point of Dirac node along $\tilde{Z}\tilde{\Gamma}$ path in Fig.~\ref{fig:6}(a), since the band gap closes between the occupied and unoccupied bands. When compressive strain is applied, the Wilson-loop spectra are well defined and the zigzag spectra appear along $\tilde{Z}\tilde{\Gamma}$ path in Fig.~\ref{fig:6}(b).

\section{CONCLUSION\label{sec:4}}

We have studied the bulk and surface electron structures of a Dirac semimetal BaHgSn. 
Though it is iso-structural to KHgSb, it has smaller lattice constant $c$. This leads to larger interlayer bonding-antibonding splitting in the Sn $p$ orbitals and results in additional band inversion at $\Gamma$ when compared with KHgSb. 
This additional band inversion leads to bulk Dirac nodes along $\Gamma$-$A$ path and further modifies the topological phases.
For KHgSb, its $Z_{2}$ indices are (0;000) and the mCN $C_{i}=2$ for the mirror plan $\overline{M}_{z}$ with $k_z=0$. In contrast, the $Z_{2}$ indices are (1;000) and $C_{i}=3$ for BaHgSn.
On (010) surface, the HSSs and doubly degenerate surface states exist along $\tilde{X}\tilde{U}$ and $\tilde{U}\tilde{Z}$ paths, which are same as KHgSb.
But along $\tilde{Z}\tilde{\Gamma}$ path, BaHgSn has projection of bulk Dirac nodes. Under compressive strain along the $y$ axis, the Dirac nodes are gapped out and there are zigzag topological surface states protected by strong $Z_2$ indices in the gap. Along $\tilde{\Gamma}\tilde{X}$ path, the topological surface bands are determined by the mCN $C_{i}$, which is two for KHgSb and three for BaHgSn. 
On (001) surface, KHgSb is trivial. For BaHgSn, there is projection of bulk Dirac cones in case without compressive strain. The nontrivial surface Dirac cone due to strong TI appears under compressive strain along the $y$ axis. 
The Fermi surfaces and QPI patterns by calculations reveal these features, such as the HSSs along $\tilde{X}\tilde{U}$ path, the zigzag surface states along $\tilde{\Gamma}\tilde{Z}$ path, and their corresponding QPI patterns. 
These results have been further clarified by the Wilson-loop spectra calculations and will help the ARPES and STM measurements to confirm them.

\begin{acknowledgments}
We acknowledge the supports from Project for Innovative Research Team of National Natural Science Foundation of China (11921004), the National Natural Science Foundation (Grant No. 11925408 and 11674369), the Ministry of Science and Technology of China (Grants No. 2016YFA0300600, 2016YFA0302400 and 2018YFA0305700), the Chinese Academy of Sciences (Grant No. XDB28000000 and XXH13506-202), the Science Challenge Project (TZ2016004), the K. C. Wong Education Foundation (GJTD-2018-01), the Beijing Natural Science Foundation (Z180008), and the Beijing Municipal Science and Technology Commission (Z181100004218001).
\end{acknowledgments}

\bibliography{BaHgSn}
\end{document}